\documentclass[showpacs,prl,onecolumn,aps,superscriptaddress,preprintnumbers,nofootinbib]{revtex4}
\usepackage[T1]{fontenc}
\usepackage[latin9]{inputenc}
\usepackage{amsmath,amssymb}
\usepackage{epsfig}
\usepackage{bm} 
\usepackage{graphicx}
\usepackage{mathrsfs}
\usepackage{amsmath}
\usepackage{amsfonts}
\usepackage{epstopdf}
\usepackage{color}
\usepackage{pifont}
\usepackage{relsize}
\usepackage{mathtools}
\def\slashchar#1{\setbox0=\hbox{$#1$}     		
   \dimen0=\wd0                                 	
   \setbox1=\hbox{/} \dimen1=\wd1               	
   \ifdim\dimen0>\dimen1                        	
      \rlap{\hbox to \dimen0{\hfil/\hfil}}      	
      #1                                        	
   \else                                        	
      \rlap{\hbox to \dimen1{\hfil$#1$\hfil}}   	
      /                                         	
   \fi}

\renewcommand{\vec}{\boldsymbol}
\newcommand{\beq}{\begin{equation}}
\newcommand{\eeq}{\end{equation}}
\newcommand{\bea}{\begin{eqnarray}}
\newcommand{\eea}{\end{eqnarray}}
\newcommand{\ba}{\begin{array}}
\newcommand{\ea}{\end{array}}

\def\eq#1{{Eq.~(\ref{#1})}}
\def\fig#1{{Fig.~\ref{#1}}}
\newcommand{\bas}{\bar{\alpha}_S}

\newcommand{\nn}{\nonumber}
\newcommand{\bg}{ \bar{\gamma}}

\newcommand{\Lb}{\left(}
\newcommand{\Rb}{\right)}
\newcommand{\h}{\frac{1}{2}}

\newcommand{\pom}{I\!\!P}

\newcommand{\intl}{\int\limits}
\begin{document}

\title{ Summing large Pomeron loops in the saturation region:  nucleus-nucleus collision.}

\author{Eugene Levin}
\email{leving@tauex.tau.ac.il}
\affiliation{Department of Particle Physics, Tel Aviv University, Tel Aviv 69978, Israel}

\date{\today}


\begin{abstract}
In this paper we found  the nucleus-nucleus scattering amplitude at high energies by summing  large Pomeron loops. It turns out that the energy dependence of this amplitude  is the same  for dipole-dipole scattering.This     happens since the evolution of one dipole cascade contributes to this scattering.

 \end{abstract}
\maketitle

\vspace{-0.5cm}
\tableofcontents

\section{Introduction}
 
 In this paper we continue \cite{LEDIDI,LEDIA}  to develop our approach for summing 
  large BFKL Pomeron loops\cite{BFKL}\footnote{BFKL stands for Balitsky, Fadin,Kuraev and Lipatov.}. Our goal is to sum these loops  for nucleus-nucleus scattering at high energies. This process gives a clear example of the dense-dense parton system scattering for which we do not have a nonlinear evolution equation. Actually, the only nonlinear evolution that we know, is able to describe the evolution of dilute-dense parton system (dipole-nucleus scattering) \cite{GLR,GLR1,MUQI,MUDI,B,K,BART,BRAUN,BRN,MV, JIMWLK1,JIMWLK2, JIMWLK3,JIMWLK4, JIMWLK5,JIMWLK6,JIMWLK7, JIMWLK8, LELU}. This nonlinear BK\cite{B,K} \footnote{BK stands for Balitsky and Kovchegov} equation  stems from summing the `fan'  Pomeron diagrams  in the BFKL Pomeron calculus\cite{GLR,BRAUN}.  In our previous paper \cite{LEDIA} we demonstrated, that  large Pomeron loops changed drastically the scattering dipole-nucleus amplitude making it asymptotically at ultra high energy the same as dipole-dipole scattering, which is an example of dilute-dilute parton system scattering.
  
\begin{figure}[ht]
    \centering
  \leavevmode
 \includegraphics[width=12cm]{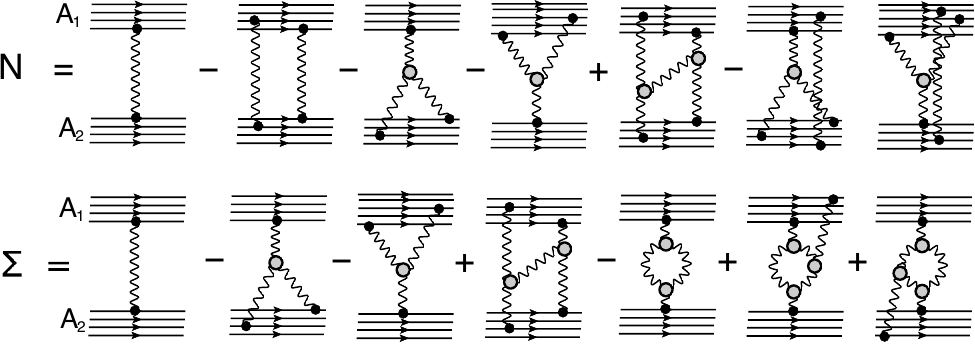}\\        
      \caption{ The first BFKL Pomeron diagrams for  nucleus-nucleus scattering. The imaginary part of the scattering amplitude (N) is equal to $ N = 1 - \exp\Lb - \Sigma\Rb$. $\Sigma$ is the set of the two nuclei irreducible Pomeron diagrams. Wavy lines denote the BFKL Pomerons. The black circles stand for Pomeron -dipole interaction vertex. The  gray circles denote the triple Pomeron vertices.}
\label{gen} 
 \end{figure}

\begin{figure}[ht]
    \centering
  \leavevmode
      \includegraphics[width=11cm]{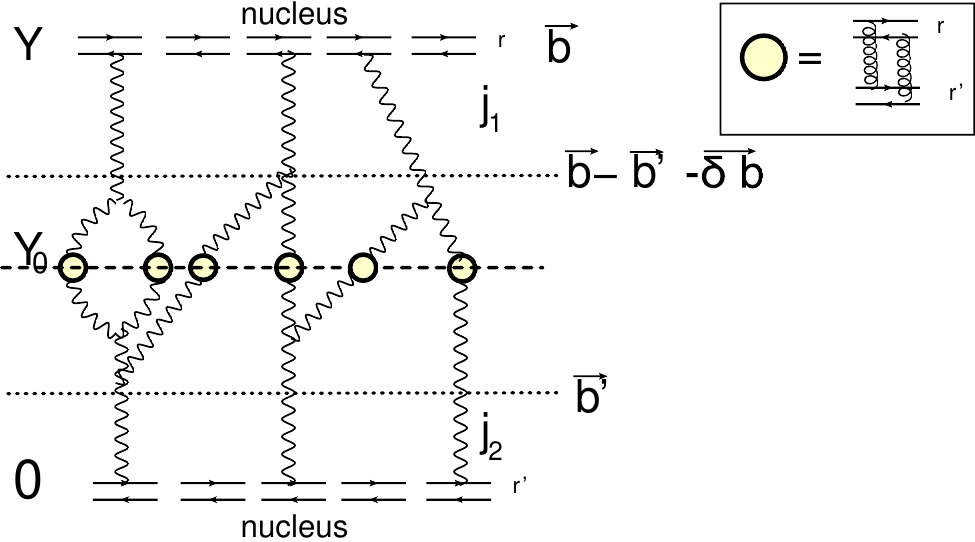}  
      \caption{Summing  large Pomeron loops for nucleus-nucleus scattering: $t$-channel unitarity for $\Sigma$. The wavy lines denote the  BFKL Pomeron exchanges.  The circles denote the amplitude $\gamma$ in the Born approximation of perturbative QCD. A nuclei  are viewed as the bags of dipoles with the size $r$ and $r'$. }
\label{mpsi}
   \end{figure}

  In \fig{gen} the first Pomeron diagrams are shown  and one can note that they have a much more complicated structure than the 'fan' diagrams. 
It is well known that 
 in spite of intensive work 
\cite{KOLEB,MUSA,LETU,LELU1,LIP,KO1,RS,KLremark2,SHXI,KOLEV,nestor,LEPRI,LMM,LEM,MUT,MUPE,IIML,LIREV,LIFT,GLR,GLR1,MUQI,MUDI,Salam,NAPE,BART,BKP,MV, KOLE,BRN,BRAUN,B,K,KOLU,JIMWLK1,JIMWLK2,JIMWLK3, JIMWLK4,JIMWLK5,JIMWLK6,JIMWLK7,JIMWLK8,AKLL,KOLU11,KOLUD,BA05,SMITH,KLW,KLLL1,KLLL2,kl,LEPR,LE1,LE2,LELU},
 the problem of summation of the Pomeron loops has  not been solved.  In our recent papers\cite{LEDIDI,LEDIA} we  have suggested the way to sum the large Pomeron loops for dipole-dipole and dipole-nucleus  scattering deeply in the saturation region. 
In our summation  we based on the $t$-channel unitarity, which has been rewritten in the convenient form for the dipole approach to CGC in Refs.\cite{MUSA,Salam,IAMU,IAMU1,KOLEB,MUDI,LELU,KO1,LE1}(see \fig{mpsi}).
 The analytic expression takes the form for the imaginary part of the scattering amplitude($N$)
       \cite{LELU,KO1,LE1}:  \bea \label{MPSI}
     && N\Lb Y, r,R ;  \vec{b}\Rb\,=\\
     &&\,\sum^\infty_{n=1}\,\Lb -1\Rb^{n+1}\,n!\int  \prod^n_i d^2 r_i\,d^2\,r'_i\,d^2 b'_i 
     \int \!\!d^2 \delta b_i\, \gamma^{BA}\Lb r_1,r'_i, \vec{b}_i -  \vec{b'_i}\equiv \delta \vec{b} _i\Rb 
    \,\,\rho^P_n\Lb Y - Y_0, \{ \vec{r}_i,\vec{b}_i\}\Rb\,\rho^T_n\Lb Y_0, \{ \vec{r}'_i,\vec{b}'_i\}\Rb \nn
      \eea
  $\gamma^{BA}$ is the scattering amplitude of two dipoles in the Born approximation of perturbative QCD.  The parton densities: $\rho^P_i Y , \{ \vec{r}_i,\vec{b}_i\}$  for projectile and $\rho^T_i Y , \{ \vec{r}_i,\vec{b}_i\}$  for target, have been introduced in Ref.\cite{LELU}  as follows:
\beq \label{PD}
\rho_n(r_1, b_1\,
\ldots\,,r_n, b_n; Y\,-\,Y_0)\,=\,\frac{1}{n!}\,\prod^n_{i =1}
\,\frac{\delta}{\delta
u_i } \,Z\left(Y\,-\,Y_0;\,[u] \right)|_{u=1}
\eeq
  where  the generating functional $Z$ is
  \beq \label{Z}
Z\Lb Y, \vec{r},\vec{b}; [u_i]\Rb\,\,=\,\,\sum^{\infty}_{n=1}\int P_n\Lb Y,\vec{r},\vec{b};\{\vec{r}_i\,\vec{b}_i\}\Rb \prod^{n}_{i=1} u\Lb \vec{r}_i\,\vec{b}_i\Rb\,d^2 r_i\,d^2 b_i
\eeq
 where $u\Lb \vec{r}_i\,\vec{b}_i\Rb \equiv\,u_i$ is an arbitrary function and $P_n$ is the probability to have $n$ dipoles with the  given kinematics.
 The initial and  boundary conditions for the BFKL cascade  stem from one dipole has 
the following form for the functional $Z$:
\begin{subequations}
\bea
Z\Lb Y=0, \vec{r},\vec{b}; [u_i]\Rb &\,\,=\,\,&u\Lb \vec{r},\vec{b}\Rb;~~~~~~~~Z\Lb Y, r,[u_i=1]\Rb = 1;\label{ZIC}\\
\rho_1\Lb Y=0,   r,b, r_1,b_1\Rb\,\,&=&\,\,\delta^{(2)}\Lb \vec{r} - \vec{r}_1\Rb \delta^{(2)}\Lb \vec{b} - \vec{b}_1\Rb ;~~~~~\rho_n\Lb 
 Y=0, \vec{r},\vec{b}; [r_i, b_i]\Rb \,=\,0 ~ \mbox{at}~\,n\geq 2;\label{ZSR}
\eea
\end{subequations}
 
The parton densities for fast dipole  and  fast nucleus have been found in Ref.\cite{LEDIDI,LEDIA}. For the completeness of presentation we will briefly discuss them in the next section. 
However we wish to stress here that they satisfy both the
 evolution equations  and  the recurrence relations derived in QCD (see Refs.\cite{LELU,LELU1,LE1}; and the analytic
 solution to the nonlinear BK equation of Ref.\cite{LETU}. Based on  parton  densities,  found in section II,   we  show  in section III that the scattering amplitude for nucleus-nucleus collisions is determined by the dipole(nucleon) -dipole(nucleon) interaction at ultra high energies. In the conclusions we summarize our results.

    \begin{boldmath}
    \section{Dipole densities $\rho_n\Lb r,b, \{r_i,b_i\}\Rb$ }

    \end{boldmath}
    
    
      \begin{boldmath}    
        \subsection{ The BFKL Pomeron  in the saturation region }     
    \end{boldmath}
    
The Green's function for the BFKL Pomeron exchange satisfy the linear equation\cite{BFKL} which has the  form:

 \beq \label{GF1}
  \frac{\partial\,G_{\pom}\Lb Y,  \vec{r}; \vec{b} \Rb }{\partial\,Y} = \bas \intl \frac{d^2 r'}{2\,\pi} \frac{r^2}{r'^2 \Lb \vec{r} -\vec{r}'\Rb^2}\,\Bigg\{ G_{\pom}\Lb Y,   \vec{r}'; \vec{b} - \h\Lb \vec{r} - \vec{r}'\Rb \Rb+G_{\pom}\Lb Y,  \vec{r}  -  \vec{r}'; \vec{b} - \h\vec{r}' \Rb -G_{\pom}\Lb Y,  \vec{r}; \vec{b} \Rb\Bigg\} \eeq
   The solution to this equation is the sum over the eigenfunction $ \phi_\gamma\Lb \vec{r} , \vec{r}_1, \vec{b}\Rb $ with the eigenvalues $\bas \chi\Lb \gamma\Rb$:
    \beq \label{GF2}   
   G_{\pom}\Lb Y,  \vec{r}, \vec{r}_1; \vec{b} \Rb  \,\,=\,\, \!\!\!\intl^{\epsilon + i \infty}_{\epsilon - i \infty} \!\!\!\frac{d \gamma}{2\,\pi\,i} e^{ \bas \chi\Lb \gamma\Rb\,Y}\, \phi_\gamma\Lb \vec{r} , \vec{r}_1, \vec{b}\Rb  \phi_{in}\Lb r_1\Rb
   \eeq
   The eigenfunction  $ \phi_\gamma\Lb \vec{r} , \vec{r}_1, \vec{b}\Rb $ (the scattering amplitude of two dipoles with sizes $r$ and $r_1$) is equal to \cite{LIP}
   \beq \label{XI}
\phi_\gamma\Lb \vec{r} , \vec{r}_1, \vec{b}\Rb\,\,\,=\,\,\,\Lb \frac{
 r^2\,r_1^2}{\Lb \vec{b}  + \h(\vec{r} - \vec{r}_1)\Rb^2\,\Lb \vec{b} 
 -  \h(\vec{r} - \vec{r}_1)\Rb^2}\Rb^\gamma =e^{\gamma\,\xi_{r,r_1} }~~\mbox{with}\,\,0 \,<\,Re\gamma\,<\,1
 \eeq

$\phi_{in}\Lb r_1\Rb $ can be found from the initial conditions: at $Y=0$ $G_{\pom}$ has to describe the scattering amplitude in the Born approximation of perturbative QCD ( the  exchange of two gluons between dipoles with sizes $r$ and $r_1$ , see insertion in \fig{mpsi}). The eigenvalue  $\omega\Lb \bas, \gamma\Rb\,=\, \bas \chi\Lb \gamma\Rb$ is given by 
\beq \label{CHI}
\omega\Lb \bas, \gamma\Rb\,\,=\,\,\bas\,\chi\Lb \gamma \Rb\,\,\,=\,\,\,\bas \Lb 2 \psi\Lb 1\Rb \,-\,\psi\Lb \gamma\Rb\,-\,\psi\Lb 1 - \gamma\Rb\Rb\eeq 
$\psi(z)$ is the Euler $\psi$ function (see Ref.\cite{RY} formulae {\bf 8.36}).

At high energies (in the saturation region) the exchange of the  BFKL Pomeron
shows two features. The new scale: saturation momentum ,   appears in the solutiomn to the  BK equation \cite{GLR}  which has the following $Y$ dependence\cite{GLR,MUT,MUPE}:
 \beq \label{QS}
 Q^2_s\Lb Y, b\Rb\,\,=\,\,Q^2_s\Lb Y=0, b\Rb \,e^{\bas\,\kappa \,Y,-\,\,\frac{3}{2\,\gamma_{cr}} \ln Y }
 \eeq 
 where $Y=0$ is the initial value of rapidity and $\kappa$ and $\gamma_{cr}$   are determined by the following equations:
  \beq \label{GACR}
\kappa \,\,\equiv\,\, \frac{\chi\Lb \gamma_{cr}\Rb}{1 - \gamma_{cr}}\,\,=\,\, - \frac{d \chi\Lb \gamma_{cr}\Rb}{d \gamma_{cr}}~
\eeq

In the saturation region the exchange of the BFKL Pomeron shows the geometric scaling behaviour\cite{GS}, being a function of one argument:
\beq \label{zz}
z\,\,=\,\,\bas \frac{\chi\Lb \bg\Rb}{\bg} \Lb Y \,-\,Y_0\Rb \,\,+\,\,\xi_{r,r_1}
\eeq 
where
$\xi_{r,r_1}$ is given by \eq{XI}. 

 It turns out that the    Green's function for the exchange of one BFKL Pomeron for scattering of dipoles with sizes $r$   and $r_1$ can written in the following way\cite{GLR,MUT}:
\beq \label{DD1}
G_{\pom}\Lb z \Rb  = N_0 e^{\bg \,z}\eeq
where $\bg = 1 - \gamma_{cr}$ , $N_0$ is a constant,  $z$ in \eq{DD1} is defined in \eq{zz}.

The Pomeron Green's function satisfies the $t$-channel unitarity\cite{BFKL}  
 analytically  continued to the $s$-channel, and can be re-written as the integration over two reggeized gluons at fixed momentum $Q_T$  that carries each Pomeron\cite{GLR,MUDI}: 

   \beq \label{TU1}  
G_{\pom}\Lb Y, Q_T,  r, R\Rb \,\,=\,\,\int\frac{ d^2 k_T}{(2\,\pi)^2}\, G_{\pom}\Lb Y - y',Q_T,  r, k_T \Rb   G_{\pom}\Lb y', Q_T, R, k_T\Rb 
\eeq
where 

\beq \label{GFMR}
 G_{\pom}\Lb Y - y', Q_T,  r, r'\Rb\,\,\,=\,\, r'^2 \int \frac{d^2 k_T}{(2\,\pi)^2} e^{i \vec{k}_T \cdot \vec{r}'}    G_{\pom}\Lb Y - y',Q_T,  r, k_T\Rb
 \eeq     
  \eq{TU1} can be re-written through $G_{\pom}\Lb Y - y', r, r', Q_T\Rb$ in the form:
  \begin{subequations}
 \bea 
G_{\pom} \Lb Y, r,R ;  \vec{b}\Rb \,\,&=&\,\,\intl d^2 r_i d^2 b_i d^2r'_i d^2 b'_i 
\rho_1\Lb Y - Y_0; r, b -b_i, r_i\Rb \gamma^{BA}\Lb r_i,r'_i,\delta b\Rb \rho_1\Lb  Y_0; r,' b'_i, r'_i\Rb\label{TU2}\\
&\,\,=\,\,&
 \,\,\frac{1}{4\,\pi^2}\int \frac{d^2 r_i}{r^4_i} \,d^2 b_i 
   \,   G_{\pom} \Lb Y - Y_0, r, r_i;  \vec{b}\ - \vec{b}_i \Rb \,\, G_{\pom} \Lb Y_0,  r_i;  \ \vec{b}_i \Rb  \label{TU21}   \eea  
    \end{subequations}   
where $ \gamma^{BA}\Lb r_i,r'_i,\delta b\Rb$ is the scattering amplitude of two dipoles in the Born approximation of perturbative QCD.
The first equation is derived in Ref.\cite{MUDI} (see also Refs.\cite{MUSA,Salam}) while  the second in Ref.\cite{CLMSOFT}.

   ~~
   ~
   ~
    \begin{boldmath}

      \subsection{ $\rho^d_n\Lb r,b, \{r_i,b_i\}\Rb$ for dipole }     
    \end{boldmath}
    

In Ref.\cite{LEDIDI} we found the parton(dipole) densities for the projectile (fast dipole with size $r$). They are equal to
\beq \label{DD5}
\rho^d_n\Lb Y - Y_0;  r,b, \{ r_i,b_i\}\Rb = C \,\intl^{\epsilon + i \infty}_{\epsilon - i \infty} \frac{d \omega}{2\,\pi\,i} e^{ \frac{ 
\bg\,\kappa}{2}\,\omega^2}\frac{1}{n!} \frac{\Gamma\Lb \omega +n\Rb}{\Gamma\Lb \omega \Rb}  \prod^n_{i=1} \frac{G_{\pom}\Lb z_i\Rb}{N_0}
\eeq
where
\beq \label{zi}
z_i \,\,=\,\,\,\,\bas \frac{\chi\Lb \bg\Rb}{\bg} \Lb Y \,-\,Y_0\Rb \,\,+\,\,\xi_{r,r_i}
\eeq
 with $\xi_{r, r_i}$ from \eq{XI} in which $r_1 $ is replaced by   $r_i$.

It has been shown that \eq{DD5} leads to the solution of  the BK nonlinear equation  deep in the saturation region\cite{LETU}.
This solution takes the form:
 \beq\label{LTSOL}
 N^{\rm DIS} \Lb z= r^2 Q^2_s\Lb Y, b\Rb\Rb\,\,=\,\,1\,\,-\,\,C(z)\exp\Lb - \frac{z^2}{2\,\kappa}\Rb
 \eeq 
 where $z = \ln\Lb r^2 Q^2_s(Y)\Rb$. One can see that  this solution
  shows the geometric scaling behaviour \cite{GS}  being a function of one variable. Function $C\Lb z\Rb$ is a smooth function  which can be considered as a constant in our approach. 
  
  Actually \eq{DD5} stems from our main idea that this solution to the BK equation has to sum the 'fan' Pomeron diagrams  
   in the BFKL Pomeron calculus. In Ref.\cite{LEDIDI} we  present \eq{LTSOL} as a sum of many Pomerons exchanges and in doing so we find the parton densities $\rho_n$.

    ~

~

    \begin{boldmath}

      \subsection{ $\rho^A_n\Lb r,b, \{r_i,b_i\}\Rb$ for  nucleus }     
    \end{boldmath}
    
 We consider a nucleus as a bag of dipoles with size $r$ which do not interact with each other at $Y_0  = 0$.
It means that (see \eq{ZSR})
\beq \label{DD41}
\rho^A_n\Lb Y_0=0, r', b',  \{ \vec{r}'_i,\vec{b}'\}\Rb \,\,=\,\,\frac{S^n_A\Lb b'\Rb}{n!}\prod^n_{i=1} \delta^{(2)}\Lb \vec{r}' - \vec{r}_i\Rb \delta^{(2)}\Lb \vec{b}' - \vec{b}_i\Rb ;
\eeq
 Factor $S^n_A\Lb b' \Rb$ describes the probability to find $n$- nucleons (dipoles) in a nucleus at the impact parameter $b'$.
 
 Looking in \fig{mpsi} one can see that parton cascades with rapidities less that $Y_0$ include only annihilation of two dipoles to one ( $2 \pom \to \pom$ vertices).  Hence each nucleon of the target  at $Y_0 =0$ creates the dipole cascade which produces $k_i$-dipoles.  The dipole densities  for a nucleus can be written as
 
\bea \label{DA1}
 &&  \rho^A_n\Lb Y_0,  r,b, \{r_i,b_i\}\Rb   \,\,=\,\, \sum^{n}_{j=1}\!\!\!\! \!\!\!\!\!\!\!\!\underbrace{\frac{1}{j!} S^j_A\Lb b\Rb}_{probability \,to\,find\,\,j-nucleons}\!\! \underbrace{\sum^{k_1=n - j +1,\dots \,,k_j=n - j +1}_{k_1=1,\dots\,,k_j=1}\,\delta_{\sum _{i=1}^j k_i = n}}_{summimg\,all\,k_i}\nn\\
&&
\rho^d_{k_1} \Lb Y_0;  r,' b', r_1, \dots\,, r_{k_1} \Rb   \times \dots \times \underbrace{\rho^d_{k_i} \Lb Y_0;  r,' b', r_{k_{i-1}+1} , \dots\,r_{k_i} \Rb}_{dipole\,density\, produced\, by \,i-th\,nucleon} \times \dots \times \rho^d_{k_j} \Lb Y_0;  r,' b', r_{k_{j-1}+1} , \dots\,r_{k_j} \Rb 
\eea  

This formula describes the simple fact that each i-th  nucleon produces $k_i$ dipoles and $\sum_{i=1}^j k_i=n$, where   $n$ is the total number of dipoles at $Y=Y_0$. Each $k_i$ runs from $k_i=1$ to $k_i = n - j +1$ with additional restriction that $\sum_{i=1}^j k_i=n$ as it is written in the sum of the second factor of \eq{DA1}.

\eq{DD41} can be easily derived from the McLerran-Venugopalan formula for the scattering amplitude \cite{MV} in which
\beq \label{DD42}
\gamma\Lb r_i,r'; b\Rb\,\,=\,\, S_A\Lb b \Rb \int\!\! d \,\delta b\,\gamma^{BA}\Lb  r_i,r'; \delta b_ i\Rb
\eeq
where $S\Lb b \Rb$ is the nucleus profile function which is equal to
\beq \label{DD43}
S_A\Lb b \Rb\,\,=\,\,\int^{\infty}_{-\infty}\!\!\!\! d z \rho_A\Lb \sqrt{b^2+z^2}\Rb 
\eeq
with $\rho_A$ is the density of nucleons in the nucleus A.

In \eq{DA1} 
it is included  that for interaction with nuclei all impact parameters  are  large 
(all $b' \to b$) for large nuclei (see \fig{mpsi}).  Summing in \eq{DA1} over $k_i$ is restricted by  $\sum _{i=1}^j k_i = n $.    Using \eq{DD5} we can obtain that
 \beq \label{DA2}
  \rho^A_n\Lb Y_0;  r,b, \{r_i,b_i\}\Rb   \,\,=\,\,\prod^n_{i=1} \frac{G_{\pom}\Lb z_i\Rb}{N_0}
 \sum^{n}_{j=1}  \frac{1}{j!} S^j_A\Lb b\Rb\!\!\!\!\!\!\!\!\!\!\!\!\!\!\!\sum^{k_1=n - j-1,\dot\,k_j=n-j-1}_{k_1=1,\dot\,k_j=1}\!\!\!\!\!\!\!\!\!\!\!\!\!\!\!\!\!\!\!\!\delta_{\sum _{i=1}^j k_i = n}\prod^{j}_{i=1} 
C \intl^{\epsilon + i \infty}_{\epsilon - i \infty} \frac{d \omega_i}{2\,\pi\,i} e^{ \frac{ 
\bg\,\kappa}{2}\,\omega_i^2}\frac{1}{k_i!} \frac{\Gamma\Lb \omega_i +k_i\Rb}{\Gamma\Lb \omega_i \Rb} \eeq

In this paper we use \eq{DA2} for  $\rho^T_n\Lb r,b, \{r_i,b_i\}\Rb$ replacing
$\Gamma\Lb \omega_i +k_i\Rb$ by the integrals over $t_i$, viz.:
 \beq \label{DA21}
 \Gamma\Lb \omega_i +k_i\Rb\,\,=\,\,\intl^\infty_0 d t_i \,t_i^{ \omega_i +k_i -1} e^{- t_i}
 \eeq  
 In doing so we obtain
 
  \bea \label{DA22}
 && \rho^A_n\Lb Y_0;  r,b, \{r_i,b_i\}\Rb   \,\,=\\
 &&\,\,\prod^n_{i=1} \frac{G_{\pom}\Lb z_i\Rb}{N_0}
 \sum^{n}_{j=1}  \frac{1}{j!} S^j_A\Lb b\Rb\!\!\!\!\!\!\!\!\!\!\!\!\!\!\!\sum^{k_1=n - j-1,\dot\,k_j=n-j-1}_{k_1=1,\dot\,k_j=1}\!\!\!\!\!\!\!\!\!\!\!\!\!\!\!\!\!\!\!\!\delta_{\sum _{i=1}^j k_i = n}S^j\Lb b\Rb  \prod^{j}_{i=1} 
C\intl^{\epsilon + i \infty}_{\epsilon - i \infty} \frac{d \omega_i}{2\,\pi\,i} e^{ \frac{ 
\bg\,\kappa}{2}\,\omega_i^2}\frac{1}{k_i!} \frac{1}{\Gamma\Lb \omega_i \Rb} 
\intl^\infty_0 d t_i t^{ \omega_i +k_i -1} e^{- t_i}\nn
\eea

In \eq{DA22} we put all   $b'_i   = b $ neglecting the size of the dipole cascade in the impact parameter space.
$z_i$ is defined in \eq{zi} for rapidity $Y_0$.
    ~
    
    \newpage
      
     \begin{boldmath}
     \section{Scattering amplitude}

      \end{boldmath}

     \begin{boldmath}
     \subsection{ Single BFKL Pomeron contribution}

      \end{boldmath}

\begin{figure}[ht]
    \centering
  \leavevmode
      \includegraphics[width=14cm]{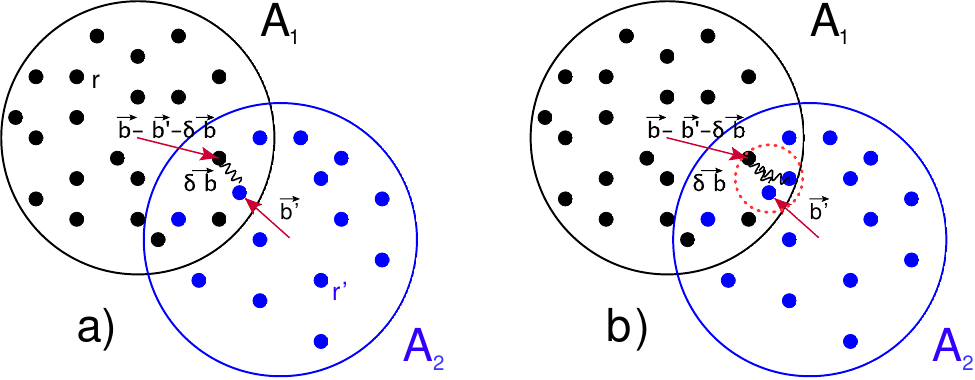}  
      \caption{The Pomeron interaction:  the impact parameter view.  \fig{1pom}-a : the exchange of one BFKL Pomeron.\fig{1pom}-b: the Pomeron interaction of one nucleon from the nucleus $A_1$ with two nucleons in the nucleus   $A_2$. The wavy lines denote the  BFKL Pomeron exchanges. The circles denote the dipoles with size $r$ or  $r'$. $\vec{b} - \vec{b}' - \vec{\delta b}  (\vec{b}') $  is the position of the nucleon in the nucleus $A_1$ ($A_2$). $\vec{\delta b}$ is the impact parameter size of the BFKL Pomeron or  several BFKL Pomerons that carry the interaction between nucleons  }
 \label{1pom}
   \end{figure}

 Let us start with the simplest interaction: the exchange of one BFKL Pomeron. The impact parameter picture of this interaction is shown in \fig{1pom}-a.
 In this figure $\vec{b}$ is the impact parameter between the centers of two nuclei .
 $\vec{b} - \vec{b}' - \vec{\delta b} $ is the position in the impact parameters of the dipole $r$ from the nucleus $A_1$,   while $\vec{b}'$ is the position of the dipole $r'$. $\vec{\delta b}$ is the position of dipole $r'$ with respect to dipole $r$ in impact parameters. The typical value of $\delta b$ is the size of the interacting dipoles which is much smaller than the sizes of nuclei: $R_{A_1}$ or $R_{A_2}$.  Hence we can replace 
 $\vec{b} - \vec{b}' - \vec{\delta b} $ by  $\vec{b} - \vec{b}' $    neglecting $\delta b$.
 In \eq{MPSI} we integrate over $\vec{\delta b}$. Therefore, we can write the contribution of the single Pomeron exchange as follows:
 \beq \label{1P1}
 N\Lb Y, r,r' ;  \vec{b}\Rb \,\,=\,\,\intl d^2 b' \,S_{A1}\Lb \vec{b} - \vec{b}'\Rb\,S_{A_2}\Lb \vec{b}'\Rb \,\intl d ^2 \delta b \,\,G_{\pom}\Lb Y, r, r';\vec{\delta b}\Rb
 \eeq
 $\int d ^2 \delta b \,\, G_{\pom}\Lb Y, r, r';\vec{\delta b}\Rb$ is the Green's function of the Pomeron at a fixed momentum $Q_T = 0$ .
 \beq \label{1P2} 
 G_{\pom}\Lb Y, Q_T=0; r, r'\Rb\,\,=\,\,\intl d ^2 \delta b\, \, G_{\pom}\Lb Y, r, r';\vec{\delta b}\Rb 
 \eeq
 
$Q_T$ is  the momentum that carries each Pomeron.   Extracting factor $2 \pi r \,r'$ from this Green's function: $ G_{\pom}\Lb Y, Q_T; r, r'\Rb\,\,=\,\,2 \,\pi\,r\,r' \tilde G\Lb Y, Q_T; r, r'\Rb$ we can rewrite the $t$-channel unitarity of \eq{TU2} in the following form for $\tilde G_{\pom}$ :
\beq \label{TU3}
 \tilde G\Lb Y, Q_T; r, r'\Rb\,\, =\,\,
\frac{1}{2\,\pi}\int \frac{d^2 r_i}{r^2_i}  
   \,  \tilde G_{\pom} \Lb Y - Y_0, Q_T;  r, r_i\Rb \,\, \tilde G_{\pom} \Lb Y_0, Q_T;  r_i ,r'   \Rb \eeq    
   
   In particular, for $Q_T=0$ we have
    
   \beq \label{TU4}
\tilde G\Lb Y, 0; r, r'\Rb\,\, =\,\,
\frac{1}{2\,\pi}\int \frac{d^2 r_i}{r^2_i}  
   \,  \tilde G_{\pom} \Lb Y - Y_0, 0;  r, r_i\Rb \,\, \tilde G_{\pom} \Lb Y_0, 0;  r_i ,r'   \Rb ;\eeq       
    
It is worthwhile mentioning that  the Green's function $ \tilde G\Lb Y, Q_T=0; r, r'\Rb$ has the form of \eq{DD1} in the saturation region with $N_0$ is dimensionless constant and  $z$ is equal to

  \beq \label{ZPR}
 z\,\,= \,\,\,\,\bas \frac{\chi\Lb \bg\Rb}{\bg} \,Y \,\,+\,\,\xi_{r,r'}    
 \eeq 
 $\xi_{r,r'}$ is equal to $\ln\Lb \frac{r^2}{r'^2}\Rb$.
 
 Plugging \eq{TU4} into \eq{1P1} we obtain:
 \bea \label{1P2}
 N\Lb Y, r,r' ;  \vec{b}\Rb \,\,&=&\,\,\intl d^2 b' \,S_{A1}\Lb \vec{b} - \vec{b}'\Rb\,S_{A_2}\Lb \vec{b}'\Rb \,\intl d ^2 \delta b \,\,G_{\pom}\Lb Y, r, r';\vec{\delta b}\Rb \nn\\
 &= & \intl d^2 b' \,S_{A1}\Lb \vec{b} - \vec{b}'\Rb\,S_{A_2}\Lb \vec{b}'\Rb \,  \frac{1}{2\,\pi}\int \frac{d^2 r_i}{r^2_i}  
   \,  \tilde G_{\pom} \Lb Y - Y_0, 0;  r, r_i\Rb \,\, \tilde G_{\pom} \Lb Y_0, 0;  r_i ,r'   \Rb\,\nn\\
   & =&\,\intl d^2 b' \frac{1}{2\,\pi}\int \frac{d^2 r_i}{r^2_i} \rho^{A_1}\Lb r, \vec{b}-\vec{b}', r_i,\vec{b}_i   =\vec{b}-\vec{b}'\Rb  \, \rho^{A_2}\Lb r,' \vec{b}', r_i,\vec{b}'_i=\vec{b}'\Rb   
    \eea 
    
      One can see that for  large nuclei we simplify the general formula of  \eq{MPSI} integrating over the impact parameters of the dipoles that  originate from the cascade of one  nucleon (dipole) in a nucleus.      
      
     \begin{boldmath}
     \subsection{ Impact parameter dependance}

      \end{boldmath}


      ~
 In this section  we are going to integrate over $\delta b_i (\delta b'_i)$. Our approach is based on the results, that have been discussed in the previous section, and we are going to use two features of the impact parameter dependance for nucleus-nucleus collisions. First is 
      the general property of \eq{MPSI} shown in \fig{mpsi}  that for $Y >Y_0$ all Pomerons interact with vertex $\pom \to 2 \pom$ while for $Y<Y_0$   only vertex $2 \pom \to \pom$ contribute to their interactions. Therefore, each nucleon (dipole) produces its own dipole cascade and these dipole cascades interact at $Y =Y_0$.
  Bearing this in mind we can rewrite  $  \rho^A_n\Lb r,b, \{r_i,b_i\}\Rb$  through the dipole cascades
  $\rho^d\Lb Y,  r,  \{r_i,\delta b_i\}\Rb $  which are  generated by $j$ nucleons:
  \beq \label{MPSI0}  
   \rho^A_n\Lb r,b, \{r_i,b_i\}\Rb  \,\,=\,\,\sum_{j=1}^{A} \frac{1}{j!} \prod_{j'=1}^{j}S_ A\Lb b_{j'}\Rb
   \rho^d_{n_{j'}}\Lb Y,  r,  \{r_i,\delta b_i\}\Rb \delta\Lb \sum_{j'=1}^ j n_{j'} =n\Rb
   \eeq    
   where $\delta b_i$ is the values of the impact parameters of dipoles with respect to the nucleons.
   The second feature is that the dipole develops a dipole cascade which has $\delta b \ll b_j$. Indeed,
  each nucleon (dipole)  is distributed over the  impact parameters in a nucleus with
      $S_A\Lb b_j\Rb$ (see \eq{DD43}). Recall that  $S_A\Lb b_j\Rb$   is  the probability  
       to have a nucleon with a large impact parameter   
    $b_j \sim R_A$. The dipole cascades, described by the Pomeron exchanges,  have  small typical $\delta b_i$ (see \fig{1pom}-b and \eq{1P1}) of the order of the dipole size: $ r \geq \delta b_i \geq 1/Q_s(Y)$. Hence we can integrate over them in \eq{MPSI} without affecting the value of $b_j$ and replace \eq{MPSI} by
   \bea\label{MPSI1}
   N\Lb Y, r,R ;  \vec{b}\Rb\,&=&\,
     \,\sum^\infty_{n=1}\,\Lb -1\Rb^{n+1}\,n!\sum^{A_1}_{j_1=1} \sum^{A_2}_{j_2=1} \int \prod d^2 b'_{j'_1}\,\frac{1}{j_1!\,j_2!} \prod^{j_1}_{j'_1=1}  \prod^{j_2}_{j'_2=1} S_{A_1}\Lb \vec{b} - \vec{b'}_{j'_1}\Rb\, S_{A_2}\Lb\vec{b_{j'_2}}'\Rb \nn\\
     &\times&
  \prod^n_i d^2 r_i\,d^2\,r'_i\,d^2 \delta b_i \,d^2 \delta b'_i \,
     \gamma^{BA}\Lb r_1,r'_i, \vec{\delta b}_i  - \vec{\delta b'_i}\Rb \nn\\
     &\times&
    \rho^d_{n_{j'_1}}\Lb Y - Y_0, r,  \{r_i,\delta b_i\}\Rb \delta\Lb \sum_{j'_1=1}^ {j_1}n_{j'_1} =n\Rb   \rho^d_{n_{j'_2}}\Lb Y_0, r',  \{r'_i,\delta b'_i\}\Rb \delta\Lb \sum_{j'_2=1}^ {j_2}n_{j'_2} =n\Rb     
   \eea 
 \eq{MPSI1} is written for $j_1 \leq j_2$. $\vec{b}'_{j'_2}$ for $j'_2>j_1$ coincide with some of $\vec{b}'_{j'_2}$ .  \fig{1pom}-b gives the example of such situation.It demonstrates the interaction for which two dipoles in the nucleus $A_2$ have almost the same impact parameters.
 
 Using \eq{TU21} we can rewrite \eq{MPSI1} in more convenient form:
 
      \bea\label{MPSI10}
   N\Lb Y, r,R ;  \vec{b}\Rb\,&=&\,
     \,\sum^\infty_{n=1}\,\Lb -1\Rb^{n+1}\,n!\sum^{A_1}_{j_1=1} \sum^{A_2}_{j_2=1} \int \prod d^2 b'_{j'_1}\,\frac{1}{j_1!\,j_2!} \prod^{j_1}_{j'_1=1}  \prod^{j_2}_{j'_2=1} S_{A_1}\Lb \vec{b} - \vec{b'}_{j'_1}\Rb\, S_{A_2}\Lb\vec{b_{j'_2}}'\Rb \\
     &\times&
  \prod^n_i \int \frac{d^2 r_i\,d^2 \delta b_i }{(2\,\pi)^2\,\,r^4_i}
    \rho^d_{n_{j'_1}}\Lb Y - Y_0, r,  \{r_i,\delta b_i\}\Rb \delta\Lb \sum_{j'_1=1}^ {j_1}n_{j'_1} =n\Rb   \rho^d_{n_{j'_2}}\Lb Y_0, r',  \{r_i,\delta b_i\}\Rb \delta\Lb \sum_{j'_2=1}^ {j_2}n_{j'_2} =n\Rb  \nn   
   \eea
  
  This equation can be simplified  by taking into account \eq{1P2} and \eq{TU4}.
Indeed, as we have discussed in the previous section, in the case of nucleus interaction the parton densities the parton cascade, generated by one nucleon (dipole)  can be described by  
 the integral over $\delta b_i$  of dipole densities,viz.:
\beq \label{MPSI11}
   \tilde \rho^A_n\Lb r,b, \{r_i\}\Rb \,\,=\,\, \int \prod_i d^2\,\delta b_i \,\rho^A_n\Lb r,b, \{r_i,b_i\}\Rb
   \eeq
   They can be written as follows:

 \beq \label{MPSI12}
\tilde  \rho^A_n\Lb r,b, \{r_i,\}\Rb   \,\,=\,\,\prod^n_{i=1} \frac{G_{\pom}\Lb z_i\Rb}{N_0}
 \sum^{n}_{j=1}  \frac{1}{j!} \Lb 2 \pi \,r^2\,S_A\Lb b\Rb\Rb^j\!\!\!\!\!\!\!\!\!\!\!\!\!\!\!\sum^{k_1=n - j-1,\dots,\,k_j=n-j-1}_{k_1=1,\dots,\,k_j=1}\!\!\!\!\!\!\!\!\!\!\!\!\!\!\!\!\!\!\!\!\delta_{\sum _{i=1}^j k_i = n} 
C \,\intl^{\epsilon + i \infty}_{\epsilon - i \infty} \frac{d \omega_i}{2\,\pi\,i} e^{ \frac{ 
\bg\,\kappa}{2}\,\omega_i^2}\frac{1}{k_i!} \frac{\Gamma\Lb \omega_i +k_i\Rb}{\Gamma\Lb \omega_i \Rb} \eeq
with 
\beq \label{MPSI13}
z_i\,\,=\,\,\bas \frac{\chi\Lb \bg\Rb}{\bg} \,Y \,\,+\,\,\ln\Lb \frac{r^2}{r^2_i} \Rb
\eeq

 For further simplification of the impact parameter dependance we consider the set of Pomeron diagrams  $\Sigma$ that are irreducible with respect two nuclei states (see \fig{gen} and compare the diagrams for the scattering amplitude and $\Sigma$).  The scattering amplitude is equal to
 
 \beq \label{MPSI4}
   N\Lb Y, r,r' ;  \vec{b}\Rb\,\,=\,\,1\,\,-\,\,\exp\Lb -  \Sigma\Lb Y, r, r' ;  \vec{b}\Rb\Rb
   \eeq
  For $\Sigma$ we can simplify \eq{MPSI1} introducing 
  \beq
  \label{MPSI15}
   \Sigma\Lb Y, r, r' ;  \vec{b}\Rb\,\,=\,\,\sum_{j_1,j_2}   \int\frac{ d^2 b' }{ 2 \pi \,r\,r'}  \Lb \pi \,r^2 \,S_{A_1}\Lb \vec{b} - \vec{b}' \Rb\Rb^{j_1}  \Lb \pi \,r'^2 \,S_{A_2}\Lb \ \vec{b}' \Rb\Rb^{j_2}   \frac{1}{j_1!\,j_2!}  \Sigma^{j_1}_{j_2}\Lb Y, r, r' \Rb 
   \eeq
  For  $\Sigma^{j_1}_{j_2}\Lb Y, r, r' \Rb$ we  have the simple form of \eq{MPSI} and \eq{MPSI1}:
    \bea
  \label{MPSI16}
   \Sigma^{j_1}_{j_2}\Lb Y, r, r' \Rb\,\,&=&\,\,\sum_{n=1}^{\infty}(-1)^{n+1}\,n!\, \prod^n_{i=1} \int d z_i \frac{G_{\pom}\Lb z_i\Rb}{N_0}
 \sum^{k_1=n - j-1,\dots,\,k_j=n-j_1-1}_{k_1=1,\dots,\,k_j=1}\!\!\!\!\!\!\!\!\!\!\!\!\!\!\!\!\!\!\!\!\delta_{\sum _{i=1}^{j_1} k_i = n} 
C \,\intl^{\epsilon + i \infty}_{\epsilon - i \infty} \frac{d \omega_i}{2\,\pi\,i} e^{ \frac{ 
\bg\,\kappa}{2}\,\omega_i^2}\frac{1}{k_i!} \frac{\Gamma\Lb \omega_i +k_i\Rb}{\Gamma\Lb \omega_i \Rb}\nn\\
& & ~~~~~~~~~~~~~~~~~~~~~~~\times \frac{G_{\pom}\Lb z'_i\Rb}{N_0}
 \sum^{l_1=n - j_2-1,\dots\,,l_j=n-j-1}_{l_1=1,\dots,\,l_j=1}\!\!\!\!\!\!\!\!\!\!\!\!\!\!\!\!\!\!\!\!\delta_{\sum _{i=1}^{j_2}l_i = n} 
C \,\intl^{\epsilon + i \infty}_{\epsilon - i \infty} \frac{d \omega_i}{2\,\pi\,i} e^{ \frac{ 
\bg\,\kappa}{2}\,\omega_i^{' 2}}\frac{1}{l_i!} \frac{\Gamma\Lb \omega_i +l_i\Rb}{\Gamma\Lb \omega_i \Rb}\eea
where $z_i$ is given by \eq{MPSI13} and $z'_i $ is equal to 
\beq \label{MPSI17}
z'_i\,\,=\,\,\bas \frac{\chi\Lb \bg\Rb}{\bg} \,\Lb Y \,-\,Y_0\Rb\,\,+\,\,\ln\Lb \frac{r^2_i}{r^{' 2}} \Rb
\eeq  
  Using \eq{TU4} we can take integral over $z_i$ and obtain:
  
    \bea
  \label{MPSI16}
   \Sigma^{j_1}_{j_2}\Lb Y, r, r' \Rb\,\,&=&\,\,\sum_{n=1}^{\infty}(-1)^{n+1}\,n!\,\Lb \frac{G_{\pom}\Lb z'\Rb}{N^2_0}\Rb^n \prod^n_{i=1} 
 \sum^{k_1=n - j-1,\dots,\,k_j=n-j_1-1}_{k_1=1,\dots,\,k_j=1}\!\!\!\!\!\!\!\!\!\!\!\!\!\!\!\!\!\!\!\!\delta_{\sum _{i=1}^{j_1} k_i = n} 
C \,\intl^{\epsilon + i \infty}_{\epsilon - i \infty} \frac{d \omega_i}{2\,\pi\,i} e^{ \frac{ 
\bg\,\kappa}{2}\,\omega_i^2}\frac{1}{k_i!} \frac{\Gamma\Lb \omega_i +k_i\Rb}{\Gamma\Lb \omega_i \Rb}\nn\\
&  & ~~~~~~~~~~~~~~~~~~~~~~~~~~~~~~~~~~~~~~~\times
 \sum^{l_1=n - j_2-1,\dot\,l_j=n-j-1}_{l_1=1,\dot\,l_j=1}\!\!\!\!\!\!\!\!\!\!\!\!\!\!\!\!\!\!\!\!\delta_{\sum _{i=1}^{j_2}l_i = n} 
C \,\intl^{\epsilon + i \infty}_{\epsilon - i \infty} \frac{d \omega'_i}{2\,\pi\,i} e^{ \frac{ 
\bg\,\kappa}{2}\,\omega_i^{' 2}}\frac{1}{l_i!} \frac{\Gamma\Lb \omega'_i +l_i\Rb}{\Gamma\Lb \omega'_i \Rb}\eea
where $z'$  is equal to 
\beq \label{MPSI17}
z'\,\,=\,\,\bas \frac{\chi\Lb \bg\Rb}{\bg} \,Y_0\,\,+\,\,\ln\Lb \frac{r^2}{r^{' 2}} \Rb
\eeq  
   ~
 
 ~

     \begin{boldmath}
     \subsection{ Contribution of two nucleons (dipoles) scattering}

      \end{boldmath}


      In this section we consider the instructive example: the interaction of pairs of nucleons(dipoles) from different nuclei. In this case 
      \beq \label{2N1}
    \Sigma^1_1\Lb Y, r, r' ;  \vec{b}\Rb   \,\,=\,\,\int\!\! \frac{d^2 b'}{ 2\,\pi}\,\,S_{A_1}\Lb \vec{b} - \vec{b}'\Rb S_{A_2} \Lb \vec{b}'\Rb \int\!\! \frac{d^2 \delta b}{ 2\,\pi}N_{dipole-dipole}\Lb Y, r, r'; \vec{\delta b}\Rb
    \eeq
       $ N_{dipole-dipole}\Lb Y, r, r';\vec{\delta b} \Rb$ has been found in Ref.\cite{LEDIDI} :
       \beq \label{2N2}
       N_{dipole-dipole}\Lb Y,r, r'; \vec{\delta b}\Rb  \,\,=\,\,1\,-\, S\Lb z'\Rb = 1\,\,-\,\,  C'^2  \exp\Lb - \frac{ z'^2}{4 \kappa}\Rb
       \eeq
       where z' is given by \eq{ZPR}.
      $C\Lb z'\Rb$ is a smooth function of $z'$.  
      
      From \eq{2N2}  we have
      \beq \label{2N3}
      \int\!\!d^2 \delta b\,\, N_{dipole- 2\,dipoles}\Lb Y, r, r'; \vec{\delta b}\Rb\,\,=\sigma_0\Lb 1\,\,-\,\,\,  C' \exp\Lb - \frac{ z'^2}{4 \kappa}\Rb\Rb\eeq
        where $\sigma_0 = 2 \pi\,r\,r'$ but  the exact numerical factor we cannot find in this general consideration.    
 
 Plugging \eq{2N3} into \eq{2N1} we obtain
  \beq \label{2N4}
    \Sigma^1_1\Lb Y, r,r' ;  \vec{b}\Rb   \,\,=\,\,\sigma_0\,\int \frac{d^2 b'}{2\,\pi}\,S_{A_1}\Lb \vec{b} - \vec{b}'\Rb S_{A_2} \Lb \vec{b}'\Rb \Bigg(1 \,\,-\,\, C  \exp\Lb - \frac{ z'^2}{4 \kappa}\Rb\Bigg)
    \eeq 
 
 Hence the scattering amplitude is equal to
 
 \bea\label{2N5}
  N\Lb Y, r,r' ;  \vec{b}\Rb\,\,&=&\,1\,\,-\,\,\exp\Lb -  \Sigma^1_1\Lb Y, r, r' ;  \vec{b}\Rb\Rb\\\
  & =&
  1\,-\,\exp\Lb -\,\sigma_0\,\int\frac{d^2 b'}{2\,\pi}S_{A_1}\Lb \vec{b} - \vec{b}'\Rb S_{A_2}\Lb \vec{b}'\Rb \Rb - \,\sigma_0  \int \frac{d^2 b'}{2\,\pi} S_{A_1}\Lb \vec{b} - \vec{b}'\Rb S_{A_2}\Lb \vec{b}'\Rb   C  \exp\Lb - \frac{ z'^2}{4 \kappa}\Rb\nn
  \eea
     \begin{boldmath}
     \subsection{$ j_1=1, j_2=2 \,   (j_1=2, j_2=1)$}

      \end{boldmath}

\begin{figure}[ht]
    \centering
  \leavevmode
      \includegraphics[width=14cm]{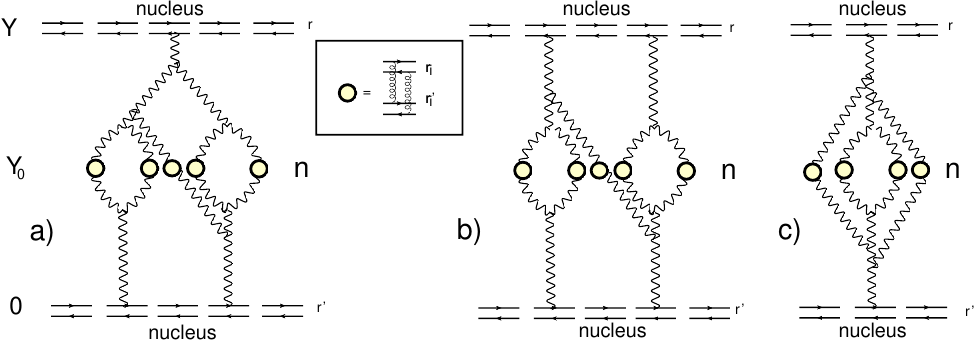}  
      \caption{ Summing  large Pomeron loops for nucleus-nucleus scattering. 
      \fig{2exampl}-a: the  parton cascade of the fast dipole from one nucleus ($j_1=1$) interacts with the  parton  cascades  generated by two  nucleons in the second nucleus ($j_2=2$) .
 \fig{2exampl}-b: the  parton cascades of two fast dipoles from one nucleus ($j_1=2$) interacts with the  parton  cascades  generated by two  nucleons in the second nucleus ($j_2=2$) .  \fig{2exampl}-c:  the  parton cascade of the fast dipole from one nucleus  interacts with the  parton  cascades  generated by one  nucleon in the second nucleus , which give the main contribution to nucleus-nucleus interaction at high energy.
 The circles denote the amplitude $\gamma$ in the Born approximation of perturbative QCD. A nucleus is viewed as the bag of dipoles with the size $r'$.}
\label{2exampl}
   \end{figure}

         In this section we consider the example of interaction: the case when the fast dipole from one nucleus interacts with two dipoles of another nucleus (see
\fig{2exampl}-a). $\Sigma$ for this case has the following form using \eq{MPSI4}:
\beq \label{12N1}
\Sigma^1_2\Lb Y, r, r' ;  \vec{b}\Rb   \,\,=\,\,\frac{1}{2!}\,\int\!\! \frac{d^2 b'}{2\,\pi}\,S_{A_1}\Lb \vec{b} - \vec{b}'\Rb 2\pi\,r'^2\,S^2_{A_2} \Lb \vec{b}'\Rb\int d \delta b \,N_{dipole- 2\,dipoles(A_2)}\Lb Y, r, r'; \vec{\delta b}\Rb
    \eeq
The scattering amplitude $N_{dipole-2\, dipoles(A_2)}\Lb Y, r, r'; \vec{\delta b}\Rb$ has been found in Ref.\cite{LEDIA} and it has the form:
\beq \label{12N2}
N_{dipole-2\, dipoles}\Lb Y, r, r'; \vec{\delta b}\Rb\,\,=\,\,  1\,\,-\,\,2\, C'  \exp\Lb - \frac{ z'^2}{4 \kappa}\Rb
\eeq
where $z'$ is given by \eq{ZPR} and $C'$ is the smooth function of $z'$.  Using this equation we obtain
 \beq \label{12N3}
 \int\!\!d^2 \delta b\,\, N_{dipole- 2\,dipoles}\Lb Y, r, r'; \vec{\delta b}\Rb\,\,=\sigma_0\Lb 1\,\,-\,\,2\,  C' \exp\Lb - \frac{ z'^2}{4 \kappa}\Rb\Rb
 \eeq 
with $\sigma_0$ which is of the order of $2 \pi r'^2$.

Plugging \eq{12N3} into \eq{12N1} we have:
\beq \label{12N4}
    \Sigma^1_2\Lb Y, r,r' ;  \vec{b}\Rb   \,\,=\,\,\frac{1}{2!}\sigma_0\,\int d^2 b'\,S_{A_1}\Lb \vec{b} - \vec{b}'\Rb \,2 \pi\,r'^2\,S^2_{A_2} \Lb \vec{b}'\Rb \Bigg(1 \,\,-\,\,2\, C  \exp\Lb - \frac{ z'^2}{4 \kappa}\Rb\Bigg)
    \eeq

   \eq{12N4} leads to the following scattering amplitude:

 \bea\label{2N4}
  &&N\Lb Y, r,r' ;  \vec{b}\Rb\,\,=\,1\,\,-\,\,\exp\Lb -  \Sigma^1_2\Lb Y, r, r' ;  \vec{b}\Rb\Rb\\
  &&
  1\,-\,\exp\Lb -\,\frac{2\pi\, r'^2}{2!}\sigma_0\int d^2 b' S_{A_1}\Lb \vec{b} - \vec{b}'\Rb S^2_{A_2}\Lb \vec{b}'\Rb \Rb -  \sigma_0\,2\pi\, r'^2  \int d^2 b' S_{A_1}\Lb \vec{b} - \vec{b}'\Rb S^2_{A_2}\Lb \vec{b}'\Rb   C'  \exp\Lb - \frac{ z'^2}{4 \kappa}\Rb\nn
  \eea
   
     \begin{boldmath}
     \subsection{$ j_1 = j_2 =2$}

      \end{boldmath}

   In this section we consider the first case of \fig{2exampl}-b which crucially differs from dipole-dipole and dipole-nucleus scattering. Using \eq{MPSI15} and \eq{MPSI16}   with $\rho_n$ from \eq{DA1} we obtain for $\Sigma^2_2$:
    \bea \label{22N1}
\tilde\Sigma^2_2\Lb Y, r,r' \Rb &=&\, C^2\intl^{\epsilon + i \infty}_{\epsilon - i \infty}\!\!\! \frac{ d \omega_1}{2\,\pi\,i} \!\!\!\intl^{\epsilon + i \infty}_{\epsilon _ i \infty}\!\!\!  \frac{d \omega_2}{2\,\pi\,i} \!\!\!\intl^{\epsilon + i \infty}_{\epsilon _ i \infty}\!\!\!  \frac{d \omega'_1}{2\,\pi\,i} \!\!\intl^{\epsilon + i \infty}_{\epsilon _ i \infty}\!\!\!  \frac{d \omega'_2}{2\,\pi\,i}
e^{ \frac{ \bg^2 \kappa}{2}\Lb  \omega_1^2 + \omega^2_2+ \omega_1^{'2} + \omega^{'2}_2\Rb}\,\sum^\infty_{n=0} \frac{\Lb - 1\Rb^n}{n!}\intl\!\!\! d z_i\, \prod^n_{i=1} 
\frac{G_{\pom}\Lb z_i\Rb}{N_0}\frac{G_{\pom}\Lb z'_i\Rb}{N_0}\,\nn\\
&&\sum_{l=1}^{n-1} \frac{n!}{l! (n -l)!}\!\!\!  \frac{ \Gamma\Lb \omega_1+l\Rb}{\Gamma\Lb \omega_1\Rb}\frac{ \Gamma\Lb \omega_2+n-l\Rb}{\Gamma\Lb \omega_2\Rb}
\sum_{k=1}^{n-1} \frac{n!}{k! (n -k)!} \frac{ \Gamma\Lb \omega'_1+k\Rb}{\Gamma\Lb \omega'_1\Rb}\frac{ \Gamma\Lb \omega'_2+n-k\Rb}{\Gamma\Lb \omega'_2\Rb}
\eea   
   Using \eq{DA21} and \eq{TU2}  we can sum over $l$ and $k$ and make the integration over $r_i$ and $r'_i$ reducing \eq{22N1} to the form:
     \bea \label{22N2}
\tilde\Sigma^2_2\Lb Y, r,r' \Rb &=& C^2\intl^{\epsilon + i \infty}_{\epsilon - i \infty}\!\!\! \frac{ d \omega_1}{2\,\pi\,i} \!\!\!\intl^{\epsilon + i \infty}_{\epsilon _ i \infty}\!\!\!  \frac{d \omega_2}{2\,\pi\,i} \!\!\!\intl^{\epsilon + i \infty}_{\epsilon _ i \infty}\!\!\!  \frac{d \omega'_1}{2\,\pi\,i} \!\!\intl^{\epsilon + i \infty}_{\epsilon _ i \infty}\!\!\!  \frac{d \omega'_2}{2\,\pi\,i}
e^{ \frac{ \bg^2 \kappa}{2}\Lb  \omega_1^2 + \omega^2_2+ \omega_1^{'2} + \omega^{'2}_2\Rb}\nn\\  
&&\sum^\infty_{n=0} \frac{\Lb - 1\Rb^n}{n!} \int^\infty_0 \!\!dt_1e^{-t_1}\,t^{\omega_1-1}_1  \int^\infty_0\!\!dt_2\,e^{-t_2} t^{\omega_2-1}_2  \int^\infty_0 \!\!dt'_1 e^{-t'_1}\,t'^{\omega'_1-1}_1  \int^\infty_0\!\!dt'_2\,e^{-t'_2}t'^{\omega'_2-1}_2  \Lb\frac{G_{\pom}\Lb z'_i\Rb}{N^2_0}\Rb^n\nn\\
&& \frac{ 1}{\Gamma\Lb \omega_1\Rb\,\Gamma\Lb \omega_2\Rb\,\Gamma\Lb \omega'_1\Rb\,\Gamma\Lb \omega'_2\Rb}\Lb \Lb t_1+t_2\Rb^n- t^n_1 - t^n_1\Rb 
\Lb \Lb t'_1+t'_2\Rb^n- t'^n_1 - t'^n_2\Rb    
\eea
             
     In this equation we introduce a different function $\tilde\Sigma^2_2\Lb Y, r,r' ;  \vec{b}\Rb$. As one sees from \eq{22N1} we started to sum with $n=0$, while in $\Sigma$ summation should go from $n=1$. since $\Sigma$ is the contribution of two nuclei  irreducible Pomeron diagrams  to the scattering amplitude. Therefore,    $  \tilde \Sigma$ is the contribution of the irreducible diagrams to the S-matrix.

Using \eq{DA21} we can integrate over $t_1$ and  $t_2$ and rewrite \eq{22N2} in the form
     \bea \label{22N3}
\tilde\Sigma^2_2\Lb Y, r,r' \Rb &=&\, C^2\intl^{\epsilon + i \infty}_{\epsilon - i \infty}\!\!\! \frac{ d \omega_1}{2\,\pi\,i} \!\!\!\intl^{\epsilon + i \infty}_{\epsilon _ i \infty}\!\!\!  \frac{d \omega_2}{2\,\pi\,i} \!\!\!\intl^{\epsilon + i \infty}_{\epsilon _ i \infty}\!\!\!  \frac{d \omega'_1}{2\,\pi\,i} \!\!\intl^{\epsilon + i \infty}_{\epsilon _ i \infty}\!\!\!  \frac{d \omega'_2}{2\,\pi\,i}
e^{ \frac{ \bg^2 \kappa}{2}\Lb  \omega_1^2 + \omega^2_2+ \omega_1^{'2} + \omega^{'2}_2\Rb}\nn\\  
&&\sum^\infty_{n=0} \frac{\Lb - 1\Rb^n}{n!}  \Lb \frac{\Gamma\Lb \omega_1 + \omega_2 + n\Rb}{\Gamma\Lb \omega_1 + \omega_2\Rb} \,\,-\,\, \frac{\Gamma\Lb \omega_1  + n\Rb}{\Gamma\Lb \omega_1\Rb }\,\,-\,\,\frac{\Gamma\Lb  \omega_2 + n\Rb}{\Gamma\Lb  \omega_2\Rb}\Rb \Lb\frac{G_{\pom}\Lb z'_i\Rb}{N^2_0}\Rb^n\nn\\
&& \int^\infty_0 \!\!dt'_1 e^{-t'_1}\,t'^{\omega'_1-1}_1  \int^\infty_0\!\!dt'_2\,e^{-t'_2}t'^{\omega'_2-1}_2  
\frac{ 1}{\Gamma\Lb \omega'_1\Rb\,\Gamma\Lb \omega'_2\Rb}\
\Lb \Lb t'_1+t'_2\Rb^n- t'^n_1 - t'^n_2\Rb    
\eea
We can perform summing over $n$ using that
\beq \label{22N4}
\sum^\infty_{n=0} \frac{\Lb - 1\Rb^n}{n!}  \frac{\Gamma\Lb \omega + n\Rb}{\Gamma\Lb \omega\Rb} \Lb\frac{G_{\pom}\Lb z'_i\Rb}{N^2_0}\Rb^n \,\,=\,\,\Lb 1 \,\,+\,\,\frac{G_{\pom}\Lb z'_i\Rb}{N^2_0}\Rb^{-\omega}\,\,\xrightarrow{\frac{G_{\pom}}{N^2_0}\,\,\gg\,1} \,\,\,\Lb \frac{G_{\pom}\Lb z'_i\Rb}{N^2_0}\Rb^{-\omega}\eeq

\eq{22N3} takes the form:
     \bea \label{22N5}
&&\tilde\Sigma^2_2\Lb Y, r,r' \Rb =\, C^2\intl^{\epsilon + i \infty}_{\epsilon - i \infty}\!\!\! \frac{ d \omega_1}{2\,\pi\,i} \!\!\!\intl^{\epsilon + i \infty}_{\epsilon _ i \infty}\!\!\!  \frac{d \omega_2}{2\,\pi\,i} \!\!\!\intl^{\epsilon + i \infty}_{\epsilon _ i \infty}\!\!\!  \frac{d \omega'_1}{2\,\pi\,i} \!\!\intl^{\epsilon + i \infty}_{\epsilon _ i \infty}\!\!\!  \frac{d \omega'_2}{2\,\pi\,i}
e^{ \frac{ \bg^2 \kappa}{2}\Lb  \omega_1^2 + \omega^2_2+ \omega_1^{'2} + \omega^{'2}_2\Rb}\nn\\  
&&\Bigg\{ \Lb 1 \,\,+\,\,\Lb t'_1+t'_2\Rb \,\frac{G_{\pom}\Lb z'_i\Rb}{N^2_0}\Rb^{-\omega_1-\omega_2} \,-\,\Lb 1 \,\,+\,\,  t'_1 \,\frac{G_{\pom}\Lb z'_i\Rb}{N^2_0}\Rb^{-\omega_1-\omega_2} \,-\,\Lb 1 \,\,+\,\,t'_2 \,\frac{G_{\pom}\Lb z'_i\Rb}{N^2_0}\Rb^{-\omega_1-\omega_2}\,\nn\\
&&
 -\,\Lb 1 
\,\,+\,\,\Lb t'_1+t'_2\Rb \,\frac{G_{\pom}\Lb z'_i\Rb}{N^2_0}\Rb^{-\omega_1}\,+\,\Lb 1 \,\,+\,\, t'_1 \,\frac{G_{\pom}\Lb z'_i\Rb}{N^2_0}\Rb^{-\omega_1} \,+\,\Lb 1 \,\,+\,\, t'_2 \,\frac{G_{\pom}\Lb z'_i\Rb}{N^2_0}\Rb^{-\omega_1}\nn\\
&& \,-\,\,\Lb 1 \,\,+\,\,\Lb  t'_1+ t'_2\Rb \,\frac{G_{\pom}\Lb z'_i\Rb}{N^2_0}\Rb^{-\omega_2}\,+\,\,\Lb 1 \,\,+\,\, t'_1 \,\frac{G_{\pom}\Lb z'_i\Rb}{N^2_0}\Rb^{-\omega_2}\,\,+\,\,\Lb 1 \,\,+\,\, t'_2 \,\frac{G_{\pom}\Lb z'_i\Rb}{N^2_0}\Rb^{-\omega_2}\, \Bigg\}\frac{1}{\Gamma\Lb \omega'_1\Rb\,\Gamma\Lb \omega'_2\Rb}
\eea

For $\frac{G_{\pom}\Lb z'_i\Rb}{N^2_0}\,\gg\,1$ we can take integrals over $t'1$ and $t'_2$. The result takes the form:

     \bea \label{22N6}
&&\tilde\Sigma^2_2\Lb Y, r,r'\Rb = C^2\intl^{\epsilon + i \infty}_{\epsilon - i \infty}\!\!\! \frac{ d \omega_1}{2\,\pi\,i} \!\!\!\intl^{\epsilon + i \infty}_{\epsilon _ i \infty}\!\!\!  \frac{d \omega_2}{2\,\pi\,i} \!\!\!\intl^{\epsilon + i \infty}_{\epsilon _ i \infty}\!\!\!  \frac{d \omega'_1}{2\,\pi\,i} \!\!\intl^{\epsilon + i \infty}_{\epsilon _ i \infty}\!\!\!  \frac{d \omega'_2}{2\,\pi\,i}
e^{ \frac{ \bg^2 \kappa}{2}\Lb  \omega_1^2 + \omega^2_2+ \omega_1^{'2} + \omega^{'2}_2\Rb}\nn\\  
&&\Bigg\{\Bigg[ \frac{\Gamma\Lb \omega'_1+\omega'_2 - \omega_1 -\omega_2\Rb}{\Gamma\Lb \omega'_1+\omega'_2 \Rb}\, -\, \frac{\Gamma\Lb \omega'_1- \omega_1 -\omega_2\Rb}{\Gamma\Lb \omega'_1 \Rb}   -\, \frac{\Gamma\Lb \omega'_2- \omega_1 -\omega_2\Rb}{\Gamma\Lb \omega'_2 \Rb} \Bigg] \exp\Lb -\Lb \omega_1 + \omega_2\Rb\, \ln\Lb \frac{G_{\pom}\Lb z'\Rb}{N_0}\Rb\Rb   \nn\\
&& \Bigg[- \frac{\Gamma\Lb \omega'_1 + \omega'_2- \omega_1\Rb}{\Gamma\Lb \omega'_1+\omega'_2 \Rb}\, +\, \frac{\Gamma\Lb \omega'_1- \omega_1\Rb }{\Gamma\Lb \omega'_1 \Rb}   +\, \frac{\Gamma\Lb \omega'_2-\omega_1\Rb}{\Gamma\Lb \omega'_2 \Rb}\Bigg] \,\exp\Lb - \omega_1 \, \ln\Lb \frac{G_{\pom}\Lb z'\Rb}{N_0}\Rb\Rb\nn\\
&& \Bigg[
 - \frac{\Gamma\Lb \omega'_1 + \omega'_2- \omega_2\Rb}{\Gamma\Lb \omega'_1+\omega'_2 \Rb}\, +\, \frac{\Gamma\Lb \omega'_1- \omega_2\Rb }{\Gamma\Lb \omega'_1 \Rb}   +\, \frac{\Gamma\Lb \omega'_2-\omega_2\Rb}{\Gamma\Lb \omega'_2 \Rb}\Bigg] \exp\Lb - \omega_2 \, \ln\Lb \frac{G_{\pom}\Lb z'\Rb}{N_0}\Rb\Rb \Bigg\}
\eea
Closing contour of integrations over $\omega_i$ and $\omega'_i$ on the  first pole of $\Gamma$-function we reduce \eq{22N6} to the sum of integrals that can be taken using the method of steepest descent at large $z'$. Finally we obtain:
   \beq \label{22N7}
\tilde\Sigma^2_2\Lb Y, r,r'\Rb =  C^2_1\exp\Lb - \frac{ z'^2}{2 \kappa}\Rb\,\,-\,\,4 C^2_2 C^2_1\exp\Lb - \frac{ z'^2}{23\kappa}\Rb    \,\,+\,\,4 C^2_3 C^2_1\exp\Lb - \frac{ z'^2}{4 \kappa}\Rb
\eeq

From this equation the contribution to $\Sigma\Lb Y, r,r' ;  \vec{b}\Rb$ with $j_1=j_2=2$ is equal to

   \bea \label{22N8}
   &&\tilde\Sigma^2_2\Lb Y, r,r' ;  \vec{b}\Rb =
\frac{1}{2!} \frac{1}{2!} \int\!\! \frac{d^2 b'}{2 \pi\,r\,r'} \Lb 2 \pi \,r^2\Rb^2 S^2_{A_1}\Lb \vec{b} - \vec{b}'\Rb \Lb 2 \pi \,r'^2\Rb^2\,S^2_{A_2}\Lb \vec{b}'\Rb\nn\\
&&\Lb 1- C^2_1\exp\Lb - \frac{ z'^2}{2 \kappa}\Rb\,\,+\,\,4 C^2_2 C^2_1\exp\Lb - \frac{ z'^2}{23\kappa}\Rb    \,\,-\,\,4 C^2_3 C^2_1\exp\Lb - \frac{ z'^2}{4 \kappa}\Rb\Rb\nn
\eea

~

~
     \begin{boldmath}
     \subsection{General case: arbitrary $j_1$ sand $j_2$}
 \end{boldmath}   
   
         ~

         ~
         
   After      thorough derivation of $j_1=j_2=2$ case we can write the general formula: 
 
  \bea \label{GN1}
&&\tilde\Sigma^{j_1}_{j_2}\Lb Y, r,r' ;\Rb =C^2 \prod_{i=1}^{j_1}\prod_{l=1}^{j_2} \intl^{\epsilon + i \infty}_{\epsilon - i \infty}\!\!\! \frac{ d \omega_i}{2\,\pi\,i} \!\!\!\intl^{\epsilon + i \infty}_{\epsilon _ i \infty}\!\!\!  \frac{d \omega'_l}{2\,\pi\,i} 
e^{ \frac{ \bg^2 \kappa}{2}\Lb   \omega_i^2 + \omega_l^{'2}\Rb }\sum_{j'_1=1}^{j_1} \frac{j_1!}{j'_1! (j_1 - j'_1)!}\sum_{j'_2=1}^{j_2} \frac{j_2!}{j'_2! (j_2 - j'_2)!} \Lb - 1\Rb^{j'_1+j'_2}\nn\\
&& \exp\Lb -\Lb \sum^{j'_1}_{i=1}   \omega_i  \Rb\, \ln\Lb \frac{G_{\pom}\Lb z'\Rb}{N_0}\Rb\Rb\,\exp\Lb - 
\Lb \sum^{j'_2}_{i=1}   \omega_i \Rb \ln \Lb\frac{G_{\pom}\Lb z'_i\Rb}{N^2_0}\Rb\Rb   \frac{\Gamma\Lb \sum^{j'_1}_{i=1}  \omega_i  \,-\,\sum^{j'_2}_{l=1}  \omega'_l\Rb}{ \Gamma\Lb \sum^{j'_2}_{l=1}  \omega'_l\Rb }\nn
\eea
Closing the contour of integration over $ \sum^{j'_1}_{i=1}   \omega_i   \equiv\,\omega$ on the pole $\omega = \sum^{j_2}_{l=1}  \omega'_l$ we have the following integrals over $\omega_i$ and $\omega'_l$:
\bea \label{GN2}
&&\tilde\Sigma^{j_1}_{j_2}\Lb Y, r,r' \Rb =\nn\\
&&\sum_{j'_1=1}^{j_1} \frac{j_1!}{j'_1! (j_1 - j'_1)!} \sum_{j'_2=1}^{j_2} \frac{j_2!}{j'_2! (j_2 - j'_2)!} ( -1)^{j'_1+j'_2}
 \, C^2   \prod_{i=2}^{j_1}\prod_{l=2}^{j_2} \intl^{\epsilon + i \infty}_{\epsilon - i \infty}\!\!\! \frac{ d \omega}{2\,\pi\,i} \exp\Lb - \omega\,\ln \Lb\frac{G_{\pom}\Lb z'\Rb}{N^2_0}\Rb\Rb \nn\\
& &  \intl^{\epsilon + i \infty}_{\epsilon - i \infty}\!\!\! \frac{ d \omega_i}{2\,\pi\,i} \!\!\!\intl^{\epsilon + i \infty}_{\epsilon _ i \infty}\!\!\!  \frac{d \omega'_l}{2\,\pi\,i} 
\exp\Lb \frac{ \bg^2 \kappa}{2} \Lb \Lb \omega-\sum^{j_2}_{l=2}\omega_i\Rb^2+ \sum^{j_1}_{i=2} \omega_i^2\,+\, \Lb \omega - \sum^{j'_2}_{l=2}\omega^{'}_l\Rb^2+
\sum^{j'_2}_{l=2} \omega_l^{' 2}\Rb\Rb
\eea
Expecting that $\omega \propto \,\bg z$ is large we can take the integrals over $\omega$'s using the method of steepest descent . The equations for the saddle points are the following:
\beq \label{SP}
(1)\,\,\, \bg^2 \kappa\Lb 2 \omega^{\mbox{\tiny SP}} - \sum_{i=2}^{j'_1}\omega^{\mbox{\tiny SP}} _i  -   \sum_{l=2}^{j'_2}\omega^{' \mbox{\tiny SP}} _l\Rb  - \bg z\,=\,0\,;~~(i) \,\,\, \omega^{\mbox{\tiny SP}}_i\,+\,\sum_{i=2}^{j'_1}\omega^{\mbox{\tiny SP}} _i\,-\,\omega^{\mbox{\tiny SP}} \,=\,0\,;~~(l) \,\,\, \omega^{\mbox{\tiny SP}}_i\,+\,\sum_{l=2}^{j'_1}\omega^{\mbox{' \tiny SP}} _l\,-\,\omega^{\mbox{\tiny SP}} \,=\,0\,;\eeq
Solutions to these equations are 
\beq \label{SP1}
\omega^{\mbox{\tiny SP}}_i  \,\,=\,\,\frac{1}{j'_1}\omega^{\mbox{\tiny SP}};~~~ 
\omega^{\mbox{' \tiny SP}}_l \,\,=\,\,\frac{1}{j'_2}\omega^{\mbox{\tiny SP}};~~~
\omega^{\mbox{\tiny SP}}\,\,=\,\,\frac{z}{\bg\,\kappa} \frac{ j'_1\,j'_2}{j'_1\,+\,j'_2}
\eeq

Plugging \eq{SP1} into \eq{GN2} we obtain that each terms in the sum of this equstion behaves as

\beq \label{GN3}
\exp\Lb - \frac{z'^{2}}{2\,\kappa} \frac{ j'_1\,j'_2}{j'_1\,+\,j'_2}\Rb
\eeq
Therefore, we can conclude that the main contribution stems from $j'_1 = j'_2 = 1$ and 
\beq \label{GN4}
\tilde\Sigma^{j_1}_{j_2}\Lb Y, r,r' \Rb\,\,=\,\,j_1\,j_2\, \tilde\Sigma^{1}_{1}\Lb Y, r,r' \Rb\,\propto\,\,\exp\Lb - \frac{z'^{2}}{4\,\kappa} \Rb
\eeq
Therefore, one can see that the only interaction between two nucleons (dipoles) from different nuclei are essential at high energy. This interaction leads 
to the nucleus-nucleus scattering amplitude deep in the saturation region which 
has the same  $z'$ behaviour as the dipole-dipole scattering. Indeed, \eq{GN2} for $j'_1=1$ and $j'_2=1$ reduces to the following expression:

\bea \label{GN5}
\tilde\Sigma^{j_1}_{j_2}\Lb Y, r,r' \Rb &=&\,j_1\,j_2 \, C^2   \prod_{i=2}^{j_1}\prod_{l=2}^{j_2}  \intl^{\epsilon + i \infty}_{\epsilon - i \infty}\!\!\! \frac{ d \omega}{2\,\pi\,i} \exp\Lb - \omega\,\ln \Lb\frac{G_{\pom}\Lb z'\Rb}{N^2_0}\Rb\Rb\nn\\
& &\intl^{\epsilon + i \infty}_{\epsilon - i \infty}\!\!\! \frac{ d \omega_i}{2\,\pi\,i} \!\!\!\intl^{\epsilon + i \infty}_{\epsilon _ i \infty}\!\!\!  \frac{d \omega'_l}{2\,\pi\,i} 
\exp\Lb \frac{ \bg^2 \kappa}{2} \Lb 2 \omega^2 + \sum^{j_1}_{i=2} \omega_i^2\,+
\sum^{j_2}_{l=2} \omega_l^{' 2}\Rb\Rb
\eea
One can see that $\omega \sim \ln   \Lb\frac{G_{\pom}\Lb z'\Rb}{N^2_0}\Rb$, while all other $\omega$ do not depend on $z'$, being of the order of $\omega_i\sim \omega'_l \sim 1/\sqrt{ \bg^2\,\kappa}$. 

Hence the scattering amplitude for the  nucleus-nucleus interaction can be reduced to the  contribution of the interaction of two nucleons from different nuclei that has been discussed in section III-C ( see \eq{2N5}). Integration over $\omega_i$ and $\omega'_l$ with $i, l \geq 2$    are absorbed in the smooth function $C$ in this equation.
       ~

   ~
   
     \begin{boldmath}
     \section{Conclusions }
      \end{boldmath}

      
      In this paper we  summed the large BFKL Pomeron loops in the framework of  the BFKL Pomeron calculus for the  nucleus-nucleus scattering. The main result is that this scattering amplitude has the same energy dependence as the dipole-dipole amplitude.  Therefore, all three types of scattering: dilute-dilute dipoles systems ( dipole-dipoles scattering \cite{LEDIDI}), dilute-dense dipole system ( dipole-nucleus scattering  \cite{LEDIA})    and dense-dense dipole system (this paper), show the same energy dependence due to contributions of  large BFKL Pomeron loops.  In other words, it turns out that the   evolution of one dipole cascade contributes to all three cases (see \fig{2exampl}-c). From \fig{2exampl}-c one can see that the interaction of two dipoles can be interpreted as the exchange of the `dressed' Pomeron with the Green's function given by \eq{2N5}.  We believe that this observation could be valuable for summing the Pomeron loops of arbitrary sizes.

    We hope that this paper will contribute to the further study of  Pomeron calculus in QCD.

    ~

    ~

       {\bf Acknowledgements} 
     
   We thank our colleagues at Tel Aviv university  for
 discussions. Special thanks go to  A. Kovner and M. Lublinsky for stimulating and encouraging discussions on the subject of this paper. 
  This research was supported  by 
BSF grant 2022132.

\end{document}